%% file: main.tex
\documentclass[letterpaper, 10 pt, conference]{ieeeconf}  

\IEEEoverridecommandlockouts                              
\usepackage{multirow}
\usepackage{graphicx}
\usepackage[utf8]{inputenc}
\usepackage[T1]{fontenc}
\usepackage{xcolor}
\usepackage{comment}
\usepackage{amssymb}
\usepackage{amsmath}
\usepackage{booktabs, multirow}
\usepackage{subcaption}
\usepackage{pifont}
\usepackage{array, makecell}
\usepackage{url}
\usepackage{mathrsfs}

\title{\LARGE \bf
RepAugment: Input-Agnostic Representation-Level Augmentation for Respiratory Sound Classification}

\author{June-Woo Kim$^{1,2}$ \,\, Miika Toikkanen$^{2}$ \,\, Sangmin Bae$^{2,3}$\,\, Minseok Kim$^{4}$$^{\dagger\ddagger}$ \,\, Ho-Young Jung$^{1}$$^{\dagger}$
\\ \{kaen2891, hoyjung\}@knu.ac.kr \quad kminseok@amazon.com
\thanks{\line(1,0){150}}
\thanks{*This research was supported by the MSIT (Ministry of Science and ICT), Korea, under the ITRC (Information Technology Research Center) support program (IITP-2024-2020-0-01808) supervised by the IITP (Institute of Information \& Communications Technology Planning \& Evaluation) and by Brian Impact Foundation, a non-profit organization dedicated to the advancement of science and technology for all.}
\thanks{$^{\dagger}$Corresponding authors \quad $^{\ddagger}$Work done before joining Amazon}  
\thanks{$^{1}$Department of Artificial Intelligence, Kyungpook National University}%
\thanks{$^{2}$RSC LAB, MODULABS \, $^{3}$KAIST AI \, $^{4}$Amazon}%
}

\begin{document}

\newcommand{\sm}[1]{{\color{blue} #1}}
\newcommand{\jw}[1]{{\color{orange} #1}}

\maketitle
\thispagestyle{empty}
\pagestyle{empty}

\begin{abstract}
Recent advancements in AI have democratized its deployment as a healthcare assistant. 
While pretrained models from large-scale visual and audio datasets have demonstrably generalized to this task, surprisingly, no studies have explored pretrained speech models, which, as human-originated sounds, intuitively would share closer resemblance to lung sounds. 
This paper explores the efficacy of pretrained speech models for respiratory sound classification. 
We find that there is a characterization gap between speech and lung sound samples, and to bridge this gap, data augmentation is essential.
However, the most widely used augmentation technique for audio and speech, SpecAugment, requires 2-dimensional spectrogram format and cannot be applied to models pretrained on speech waveforms.
To address this, we propose RepAugment, an input-agnostic representation-level augmentation technique that outperforms SpecAugment, but is also suitable for respiratory sound classification with waveform pretrained models. 
Experimental results show that our approach outperforms the SpecAugment, demonstrating a substantial improvement in the accuracy of minority disease classes, reaching up to 7.14\%.
\end{abstract}

\input{1-introduction}

\input{2-relatedworks}
\input{3-method}
\input{4-experiments}
\input{5-conclusion}

\clearpage

\bibliography{ref}
\bibliographystyle{IEEEtran}

\end{document}

%% file: 1-introduction.tex
\section{INTRODUCTION}

Recent remarkable advances in machine learning are rapidly transforming the landscape of healthcare, offering unprecedented opportunities for early disease detection and personalized treatment pathways. 
Within this evolving paradigm, respiratory sound classification has gained significant attention due to its potential to non-intrusively identify lung diseases based on respiratory sound.
However, the high cost of collecting lung sound data has been an obstacle to the wide adoption of respiratory sound classifier models because a substantial amount of data is typically required to train a sufficiently reliable machine learning model.


To cope with the respiratory data deficiency, previous studies~\cite{bae23b_interspeech, kim2023stethoscope, kim2023adversarial, moummad2023pretraining, wang2022domain} for respiratory sound classification have usually exploited the following two-staged approach: $(i)$ \emph{pretraining} with sufficient data from other domains\,(e.g., image or audio data) to enhance general feature representation capability, 
and $(ii)$ then \emph{fine-tuning the pretrained model with data augmentation} on respiratory sound to avoid
the overfitting issues caused by lack of diverse data.
The state-of-the-art approach~\cite{bae23b_interspeech} for respiratory sound classification demonstrated impressive performance by exploiting Audio Spectrogram Transformer (AST)~\cite{gong2021ast} pretrained on audio and image datasets, and fine-tuning with the help of SpecAugment~\cite{park2019specaugment} that augments respiratory sound samples in the format of spectrogram images.



Previous successful approaches to respiratory sound classification have been limited to image-based models where the audio data is transformed into spectrograms.
Nevertheless, such transformation can lead to considerable loss of information, so end-to-end learning that directly leverages the dense and rich information of the original waveforms has gained significant attention in recent speech domain studies~\cite{zeghidour2018end, baevski2020wav2vec, hsu2021hubert, babu2021xls}.
With the same motivation,
our research goal is to investigate the lung sound classification performance by directly using the original waveforms. Instead of models pretrained on images, we first explore speech-based pretrained encoders~\cite{baevski2020wav2vec, hsu2021hubert, babu2021xls} that focus on learning speech representations 
while having access to more rich
information present in the waveform.

With the change in the input domain, we can no longer apply SpecAugment~\cite{park2019specaugment}, which is limited to image formats (i.e., 2D spectrogram), to fine-tune the models.
In this regard, there is a call for an \emph{input-agnostic} approach that can be applied universally to the strong capability of pretrained models with \emph{any} input type for respiratory sound classification.
In this paper, we propose \emph{RepAugment}, a novel and efficient input-agnostic augmentation technique applied at the representation level. The proposed RepAugment operates directly on the model's output representations 
before classifier,
rendering it input-agnostic and applicable to a wide range of model architectures.
In a synergistic effect between representation masking and noise generation, RepAugment reduces the reliance on particular features, while expanding the model's understanding of the under-represented classes.

Experimental results demonstrate the effectiveness of RepAugment, specifically when applied to models pretrained with audio or speech datasets, outperforming the conventional SpecAugment.
Moreover, our approach also shows accuracy gains up to 7.14\% for the most under-represented classes, highlighting its potential to improve the diagnostics of abnormal 
cases.
Our main contributions are summarized as follows:
\begin{itemize}
\item We explore the capability of pretrained speech models, 
and compare them to the models pretrained on image and audio datasets commonly used for respiratory sound classification.
\item We propose RepAugment, a simple and novel input-agnostic representation-level augmentation method that can leverage the strong capability of pretrained models with any input type for respiratory sound classification.
\item 
We demonstrate the effectiveness of RepAugment, particularly for improving the performance of minority categories, when fine-tuning respiratory samples under several pretraining conditions.
\end{itemize}



%% file: 2-relatedworks.tex
\section{RELATED WORKS}

\subsection{Pretrained Models}
Numerous deep learning methods have emerged for the lung sound classification task, particularly those based on CNN, such as ResNet~\cite{he2016deep} and EfficientNet~\cite{tan2019efficientnet}. 
Moreover, prior work has predominantly
used large-scale visual and audio datasets like ImageNet~\cite{deng2009imagenet} and AudioSet~\cite{audioset}, and their mixtures~\cite{gong2021ast}, which have been shown to notably improve the performance of models on the ICBHI~\cite{rocha2018alpha} dataset. 
Specifically, Bae\,\textit{et al}.~\cite{bae23b_interspeech} conducted a comprehensive study comparing the performance of different models on respiratory sound classification with and without using pretrained weights from ImageNet and AudioSet.
This comparison revealed notable differences in effectiveness across models, highlighting the generalizability of pretrained weights 
over full training in the case of respiratory sound classification tasks.

Unlike previous works, our goal is to 
tap into the ability of such large models' that have been pretrained on massive speech data~\cite{panayotov2015librispeech, kahn2020libri, pratap20_interspeech}, and 
have so far remained unused in respiratory sound classification tasks.

\subsection{Data Augmentation Strategy}

Data augmentations aim to maintain the class identity of a sample while varying other aspects such as style to improve generalization. They are related to domain randomization~\cite{tobin2017domain}, which involves the random alteration of samples while ensuring that they remain closely aligned with the underlying data manifold. Mixup~\cite{zhang2017mixup} provides variety of benefits for image classification by applying such operations on data, while Manifold Mixup~\cite{verma2019manifold} achieves the same on the hidden representations of the neural network. The simple approach of perturbing samples with Gaussian noise has been found effective at domain generalization~\cite{li2021simple, zhou2021domain}, few shot classification~\cite{tseng2020cross}, and addressing class imbalance by using variance in head classes to augment tail classes~\cite{liu2020deep}.

In the fields of speech recognition, audio classification, and respiratory sound classification, SpecAugment \cite{park2019specaugment} is a popular data-augmentation technique that directly operates on the spectrograms, masking out continuous blocks of time-steps and frequency bins. 
%
However, this does not apply to waveform data used by the state-of-the-art models~\cite{baevski2020wav2vec, hsu2021hubert, babu2021xls}, so there is a call for input-agnostic augmentation.
To this end, we apply a similar strategy on the representation level to remain independent from the input modality. 
Masking the features may be considered analogous to dropout~\cite{srivastava2014dropout}, which is not a data augmentation technique, but can also be viewed as kind of noising data-augmentation 
~\cite{bouthillier2015dropout}. We augment the features directly by masking out values from the feature representations, while dropout randomly drops neurons from the network.

%% file: 3-method.tex
\section{METHOD: REPAUGMENT}
\subsection{Data Augmentation Policy}
Our objective is to introduce an augmentation policy that operates on the model's output representations 
before classifier, making it independent of input specifics and adaptable to a diverse array of model architectures.
Our \emph{RepAugment} strategy, denoted as $A$, is based on two operations.
Masking augmentation \emph{Rep-Mask}, denoted as $A_{mask}$, and noise generation augmentation \emph{Rep-Gen}, denoted as $A_{gen}$.

Rep-Mask operation attempts to reduce the model's reliance on particular features by restricting access to them randomly during training. 
This is done by masking the feature $z \in \mathbb{R}^{d}$ using a \emph{mask} $M$ consisting of values 0 and 1, where 1 indicates the masked and 0 denotes the unmasked part.
Subsequently, we replace the masked parts with the vector $\overline{z}$, which is the mean of $z$ computed over all dimensions and broadcasted back into the shape of $z$. This approach was found to work better than using zero values.
\begin{equation}
    A_{mask}(z) = z \odot (1 - M) + \overline{z} \odot M
\end{equation}
Here $\odot$ denotes an elementwise multiplication.


Given a feature $z$, in order to generate the masking index of $M$, we apply $k$ number of mask bands to the feature with different ranges. 
We set the masking ranges to $[i$, $i + j)$ positions in $z$. The start index $i$ is chosen from $[0, d - j)$, and $j$ is chosen from $[0, L)$. We empirically used $k=2$ and $L=288$ in our experiments. 
Masking consecutive features channels is not crucial in experiments of this paper as they are spatially independent, but becomes more important when applied to 2D features that include the time-axis.

Reg-Gen operation attempts to expand the model's experience of the underrepresented classes by adding random variation to them. To achieve this, we perturb the minority classes of $c_{crackle}$, $c_{wheeze}$, and $c_{both}$ with random noise as follows. 

\begin{equation}
    A_{gen}(z) = z + z_{noise}
\end{equation}
We generate a Gaussian noise tensor $z_{noise} \sim N(\mu, \sigma^2)$ and add it to the original feature $z$ encoded from input $x$ corresponding to class $c$. We used the standard normal distribution, where $\mu=0$ and $\sigma=1$.

The full augmentation A
then becomes the following equation, where the samples from the normal class $c_{normal}$, only receive the Rep-Mask augmentation.

\begin{equation}
A =
\begin{cases}
    A_{mask}(z), & \text{if } c = c_{normal} \\
    A_{gen}(A_{mask}(z)), & \text{else} \\
\end{cases}
\end{equation}
The masked values also get perturbed by Rep-Gen thereby also reducing the unnatural appearance of masking with constant values.

\begin{figure*}[ht!]
    \centering
    \includegraphics[width=1.0\linewidth]{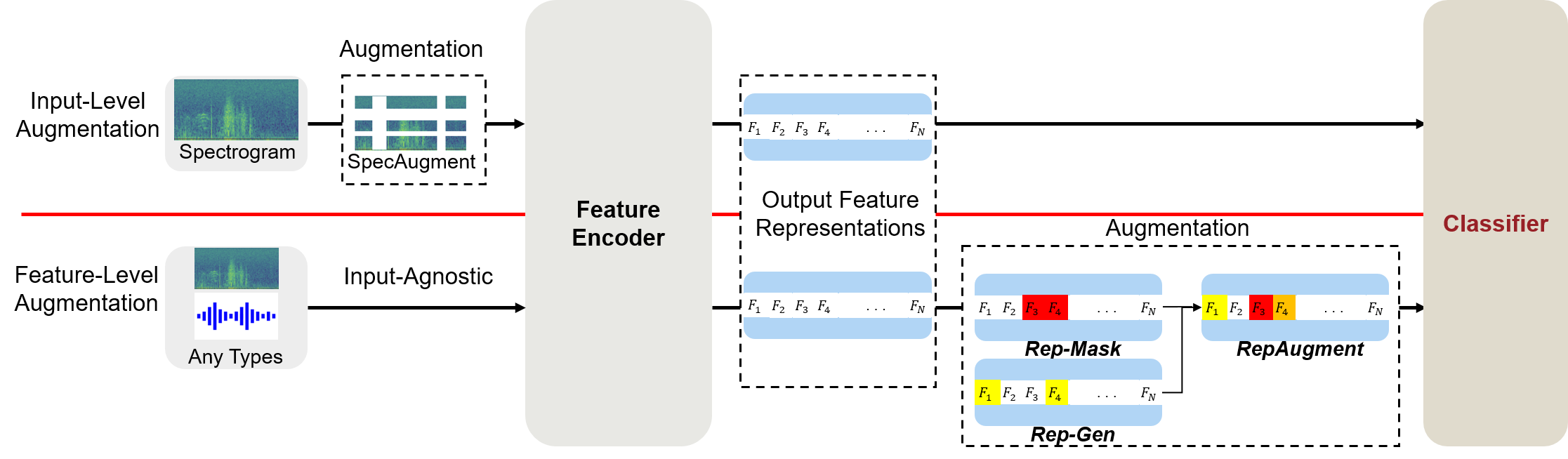}
    \vspace{-3mm}
    \caption{Difference between the input-level and feature-level augmentation for respiratory sound classification. Our proposed input-agnostic \emph{RepAugment} which consists of Rep-Mask and Rep-Gen (at bottom right) can be employed for any input type, whereas the SpecAugment (at top left) can only be applied to the input spectrogram.}
    \label{fig:proposed}
\end{figure*}
\subsection{Model Architectures}
\vspace{2mm}
\subsubsection{Pretrained Speech Models}
We first explore the capability of pretrained speech models for respiratory sound classification tasks. 
In this regard, we use various pretrained models as our backbone: wav2vec2-base~\cite{baevski2020wav2vec} pretrained on the LibriSpeech 960 hours dataset~\cite{panayotov2015librispeech}, wav2vec2-large on the LibriLight 60k hours dataset~\cite{kahn2020libri}, HuBERT-base~\cite{hsu2021hubert} on LibriSpeech, HuBERT-large on LibriLight, and XLS-R-300~\cite{babu2021xls} on Multilingual LibriSpeech~\cite{pratap20_interspeech} combined with other speech datasets as described in~\cite{babu2021xls}, amounting to a cumulative duration of approximately 436k hours. 
To fine-tune these models for the respiratory sound classification tasks, we take the last hidden states of each pretrained model as input for the lung sound classifier.

\vspace{2mm}
\subsubsection{Pretrained Image and Audio Models}
To apply the proposed input-agnostic RepAugment to the fine-tuning process for various input types, we further use diverse pretrained models on large-scale visual and audio datasets such as ImageNet~\cite{deng2009imagenet} and AudioSet~\cite{audioset}, including their combinations.
To this end, we use pretrained ResNet~\cite{he2016deep} and EfficientNet~\cite{tan2019efficientnet} models on ImageNet, along with CNN6~\cite{kong2020panns} on AudioSet, and also leverage the Audio Spectrogram Transformer (AST)~\cite{gong2021ast} for both ImageNet and AudioSet.

\vspace{2mm}
\subsubsection{Applying RepAugment}
The proposed RepAugment is applied to the feature representations where they are obtained from feature encoders.
In detail, for both pretrained speech and image as well as audio models, we aggregate the last hidden states from each pretrained model by averaging their values across temporal information.
We have empirically found that using average pooling provides a slight performance improvement than employing mean values of feature representations except for the AST model. 
Note that we use the average values of \emph{class} and \emph{distill} tokens in the AST model.
These aggregated feature representations are then augmented with RepAugment and fed into the lung sound classifier consisting of a 4-dimensional linear layer with LayerNorm~\cite{ba2016layer}.


%% file: 4-experiments.tex
\section{EXPERIMENTS}
\subsection{Dataset}
We used the ICBHI~\cite{rocha2018alpha} dataset that serves as a popular open benchmark for the respiratory sound classification task. 
This dataset has an official split into training and test sets, maintaining the ratios of 60--40\% for the training and test sets respectively, with no patient overlap between them. 
Consequently, the training and test sets comprised 4,142 and 2,756 samples respectively, and are categorized into four classes: \emph{normal} (49.8\%--57.29\%), \emph{crackle} (29.30\%--23.55\%), \emph{wheeze} (12.10\%--13.97\%), and \emph{both} (8.80\%--5.19\%). 
For the performance metrics, we used the \emph{Specificity}, \emph{Sensitivity} and their arithmetic mean, hereinafter denoted as $S_{p}$, $S_{e}$, and $Score$, following the definitions outlined in~\cite{rocha2018alpha}.

\subsection{Experimental Setting}
For training, we standardized the duration of all samples to 8 seconds.
In the case of pretrained speech models, we used the raw waveform of lung sound samples as model input (i.e., without normalization).
The pretrained weights are from the S3PRL toolkit~\cite{yang21c_interspeech}.
For the models pretrained on image and audio, we used the 128-dimensional log Mel filterbank features with a window size of 25ms and an overlap size of 10ms. The log Mel filterbank features were normalized with zero mean unit variance, except for the AST, 
for which we used the mean and standard deviation of --4.27 and 4.57, as described in~\cite{bae23b_interspeech}.
For the ResNet, EfficientNet, and CNN6, we used the Adam optimizer with a learning rate of 1e--3 and a batch size of 64 for 400 epochs. For the AST and all pretrained speech models, we adopted the same optimizer with a learning rate of 5e--5 and a batch size of 8 for 50 epochs. 
To ensure result stability, we conducted the model training with a consistent set of five random seeds across all experiments. 
Consequently, the reported values for $S_p$, $S_e$, and $Score$ include both the mean and variance.

\subsection{Experimental Results}

\input{Tables/table1}

\vspace{2mm}
\subsubsection{Using Pretrained Speech Models}
Table~\ref{tab:table1} provides an overview of the respiratory classification performance obtained with various pretrained models and datasets. 
To ascertain the impact of architecture and pretraining datasets, we conducted fine-tuning on the models without other additional learning techniques.
The most performant pretrained speech model is XLS-R-300M, with the ICBHI Score of 56.29\%.
Despite employing robust pretrained speech models, the performance in respiratory sound classification fell short of the initial expectation of at least being comparable with AST at Score of 59.14\%.
We hypothesize that the lack of architecture-specific hyperparameter optimization for AST and other pretrained speech models may contribute to this as well as data distribution inconsistency between lung sound and human speech.
However, our result reveals that the performance scales with the pretrained speech model size, as is evident from the 310M large model outperforming the 95M base model in both wav2vec2 and HuBERT architectures.

\input{Tables/table2}
\vspace{2mm}
\subsubsection{Effectiveness of RepAugment}
As shown in Table~\ref{tab:table1}, SpecAugment 
was the most useful for models pretrained on ImageNet, such as ResNet, EfficientNet, whereas our proposed RepAugment 
demonstrated performance gains for all models, and in the case of CNN6 pretrained on AudioSet as well as AST pretrained on ImageNet and AudioSet, it is the strongest method.
Moreover, RepAugment also achieved the best performance on all pretrained speech models, regardless of the datasets on which 
they were pretrained.
These findings not only highlight the versatility of applying our approach across various model input formats but also show the effectiveness of the proposed RepAugment 
for pretrained models both on audio and speech datasets.

\vspace{2mm}
\subsubsection{Comprehensive Comparison}

Table~\ref{tab:table2} shows an overall comparison of different methods for lung sound classification on the ICBHI dataset. 
Scores of 56.73\% and 61.51\% were achieved with our proposed method applied to XLS-R-300M and AST, respectively.
We would like to highlight that the proposed RepAugment on the AST achieved a remarkable improvement of 2.37\% over the Score of AST with no augmentation, and outperformed the Score of AST with SpecAugment by an improvement of 1.96\% as well.
Furthermore, results show that our method achieves state-of-the-art performance, particularly in terms of $S_p$ with XLS-R-300M and $S_e$ with AST, respectively.
We believe that our method can be helpful 
when fine-tuning with various other datasets.

\input{Tables/table3}

\begin{figure*}[!t]
    \vspace{-1mm}
    \centering
    \begin{subfigure}{.4\linewidth}
      \centering
      \includegraphics[width=1.0\linewidth]{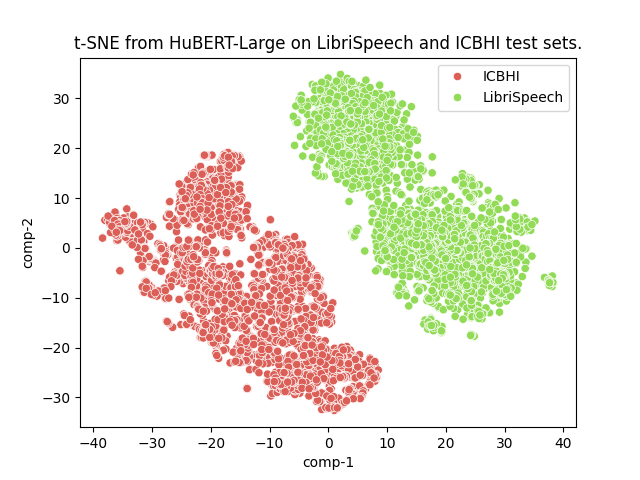}
      \caption{t-SNE results of pretrained HuBERT-Large}
      \label{fig:sfig1}
    \end{subfigure}%
    \centering
    \begin{subfigure}{.4\linewidth}
      \centering
      \includegraphics[width=1.0\linewidth]{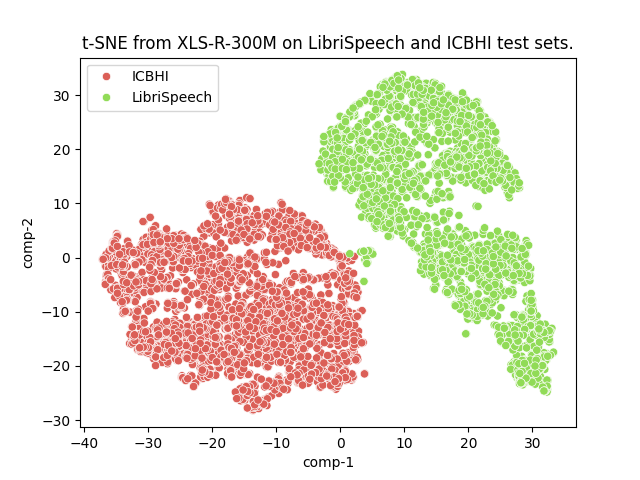}
      \caption{t-SNE results of pretrained XLS-R-300M)}
      \label{fig:sfig2}
    \end{subfigure}    
    \caption{t-SNE results of pretrained HuBERT-Large and XLS-R-300M on LibriSpeech and ICBHI test sets.}
    \label{fig:t-sne}
    \vspace{-1mm}
    \end{figure*}

\vspace{2mm}
\subsubsection{Effectiveness of Masking and Noise Generation}
To validate the effectiveness of RepAugment, we study how each augmentation strategy impacts each class of lung sounds.
As shown in Table~\ref{tab:table3}, \emph{Rep-Mask} appears to benefit the more represented classes best and contributes to the overall accuracy. In fact, in the case of XLS-R, small accuracy degradation occurs for the smallest category $c_{both}$. In the case of AST where SpecAugment is available, it exhibits a similar pattern as Rep-Mask, where improvement in the largest class can be observed.
\emph{Rep-Gen} on the other hand contributes to generalization of the minority classes, and we see the opposite pattern where the accuracy on the larger category $c_{crackle}$ remains similar to the baseline.
Finally, \emph{RepAugment} combines Rep-Mask and Rep-Gen in a synergistic effect that leads to improvements in all categories without regard to number of samples, but especially in the smallest category.

\vspace{2mm}
\subsubsection{Distribution Gap between Speech and Respiratory Samples} 
To investigate the reasons behind the lower performance of pretrained speech models compared to models pretrained on image and audio datasets, we analyze the data distribution between speech and respiratory samples. 
Figure~\ref{fig:t-sne} provides t-SNE results generated from the output of the XLS-R-300M encoder that performed best among the pretrained speech models in our experiments using the LibriSpeech and ICBHI test sets.
Notably, the red and green samples exhibit a distinct separation, indicating that the models, which are primarily trained on human speech data, significantly differ from respiratory sounds.
This domain gap can lead to challenges in transferring learned knowledge and achieving optimal performance in lung sound classification tasks. Consequently, further studies are necessary to enhance the
applicability of pretrained speech models for this task.

%% file: Tables/table1.tex
\begin{table*}[!ht]
    \vspace{-5pt}
    \centering
    \caption{
    Comparison of respiratory sound classification performance under different pretrained models and augmentation configurations.
    \textbf{Bold} denotes the best result.}\label{tab:table1}
    \vspace{-1mm}
    \renewcommand{\arraystretch}{1}
    \addtolength{\tabcolsep}{8pt}
    \resizebox{\linewidth}{!}{
    \begin{tabular}{llccl|ccc}

    \toprule
    \multirow{17}{*}{\rotatebox[origin=c]{90}{\textbf{pretrained on image \& audio}}}  & Model & \# Params & Pretrain & Augmentation &  $S_p$\,(\%) & $S_e$\,(\%) & \textbf{Score}\,(\%) \\
    \hline \midrule

     & \multirow{3}{*}{\multirow{3}{*}\textbf{ResNet}\, \cite{he2016deep}}  & \multirow{3}{*}{\multirow{3}{*}\textbf{11.7M}} & \multirow{3}{*}{\multirow{3}{*}\textbf{ImageNet}} & No Aug.  & $\text{{76.25}}_{\pm 5.65}$ & $\text{{28.87}}_{\pm 5.59}$  & $\text{{52.57}}_{\pm 1.38}$ \\
     & & & & SpecAugment & $\text{{{76.70}}}_{\pm 5.65}$ & $\text{{{33.47}}}_{\pm 4.03}$ & $\text{\textbf{{55.09}}}_{\pm 0.82}$ \\
     & & & & \emph{RepAugment} & $\text{{75.10}}_{\pm 7.63}$ & $\text{{31.63}}_{\pm 8.75}$ & $\text{{53.37}}_{\pm 1.37}$ \\
     \cmidrule{2-8}

     & \multirow{3}{*}{\multirow{3}{*}\textbf{EfficientNet}\, \cite{tan2019efficientnet}}  & \multirow{3}{*}{\multirow{3}{*}\textbf{5.3M}} & \multirow{3}{*}{\multirow{3}{*}\textbf{ImageNet}} & No Aug.  & $\text{{{79.37}}}_{\pm 3.58}$ & $\text{{{32.67}}}_{\pm 3.26}$  & $\text{{{56.02}}}_{\pm 0.52}$ \\
     & & & & SpecAugment & $\text{{{78.21}}}_{\pm 3.56}$ & ${{{34.44}}}_{\pm 2.84}$ & $\text{\textbf{{56.33}}}_{\pm 0.43}$ \\
     & & & & \emph{RepAugment} & $\text{{{75.10}}}_{\pm 7.63}$ & $\text{{{37.23}}}_{\pm 7.41}$ & $\text{{{56.17}}}_{\pm 0.65}$ \\
     \cmidrule{2-8}

     & \multirow{3}{*}{\multirow{3}{*}\textbf{CNN6}\, \cite{kong2020panns}}  & \multirow{3}{*}{\multirow{3}{*}\textbf{4.8M}} & \multirow{3}{*}{\multirow{3}{*}\textbf{AudioSet}} & No Aug.  & $\text{{{72.83}}}_{\pm 4.61}$ & $\text{{{39.64}}}_{\pm 3.6}$  & $\text{{{56.62}}}_{\pm 1.25}$ \\
     & & & & SpecAugment & $\text{{{77.00}}}_{\pm 3.27}$ & $\text{{{37.35}}}_{\pm 3.15}$ & $\text{{{57.17}}}_{\pm 0.81}$ \\
     & & & & \emph{RepAugment} & $\text{{{75.93}}}_{\pm 4.88}$ & $\text{{{38.95}}}_{\pm 6.41}$ & $\text{\textbf{{57.44}}}_{\pm 1.64}$ \\
     \cmidrule{2-8}

     & \multirow{3}{*}{\multirow{3}{*}\textbf{AST}\, \cite{bae23b_interspeech, gong2021ast}}  & \multirow{3}{*}{\multirow{3}{*}\textbf{87.7M}} & \multirow{3}{*}{\makecell{ImageNet\,+\\\!AudioSet}}  & No Aug.  & $\text{{{76.57}}}_{\pm 2.41}$ & $\text{{{41.72}}}_{\pm 1.26}$  & $\text{{{59.14}}}_{\pm 1.66}$ \\
     & & & & SpecAugment & $\text{{{77.14.}}}_{\pm 3.35}$ & $\text{{{43.07}}}_{\pm 2.80}$ & $\text{{{59.55}}}_{\pm 0.88}$ \\
     & & & & \emph{RepAugment} & $\text{{{82.47}}}_{\pm 5.39}$ & $\text{{{40.55}}}_{\pm 4.70}$ & $\text{\textbf{{61.51}}}_{\pm 0.51}$ \\

    \hline \midrule

     \multirow{12.5}{*}{\rotatebox[origin=c]{90}{\textbf{pretrained on speech}}}  & \multirow{2}{*}{\multirow{2}{*}\textbf{wav2vec2-Base}\, \cite{baevski2020wav2vec}}  & \multirow{2}{*}{\multirow{2}{*}\textbf{95M}} & \multirow{2}{*}{\multirow{2}{*}\textbf{LibriSpeech}} & No Aug.  & $\text{{{71.98}}}_{\pm 7.74}$ & $\text{{{30.86}}}_{\pm 9.10}$  & $\text{{{51.42}}}_{\pm 1.66}$ \\
     & & & & \emph{RepAugment} & $\text{{{68.71}}}_{\pm 14.25}$ & $\text{{{35.07}}}_{\pm 14.2}$ & $\text{{\textbf{51.89}}}_{\pm 0.44}$ \\
     \cmidrule{2-8}

     & \multirow{2}{*}{\multirow{2}{*}\textbf{wav2vec2-Large}\, \cite{baevski2020wav2vec}}  & \multirow{2}{*}{\multirow{2}{*}\textbf{310M}} & \multirow{2}{*}{\multirow{2}{*}\textbf{LibriLight}} & No Aug.  & $\text{{{68.61}}}_{\pm 2.02}$ & $\text{{{41.09}}}_{\pm 2.94}$  & $\text{{{54.85}}}_{\pm 0.65}$ \\
     & & & & \emph{RepAugment} & $\text{{62.80}}_{\pm 2.59}$ & $\text{{{47.51}}}_{\pm 1.51}$ & $\text{{\textbf{55.16}}}_{\pm 1.03}$ \\
     \cmidrule{2-8}

     & \multirow{2}{*}{\multirow{2}{*}\textbf{HuBERT-Base}\, \cite{hsu2021hubert}} & \multirow{2}{*}{\multirow{2}{*}\textbf{94M}} & \multirow{2}{*}{\multirow{2}{*}\textbf{LibriSpeech}} & No Aug.  & $\text{{{68.08}}}_{\pm 8.50}$ & $\text{{{38.37}}}_{\pm 8.72}$  & $\text{{{51.22}}}_{\pm 0.31}$ \\
     & & & & \emph{RepAugment} & $\text{{{64.80}}}_{\pm 2.51}$ & $\text{{{42.26}}}_{\pm 3.73}$ & $\text{{\textbf{53.53}}}_{\pm 1.72}$ \\
     \cmidrule{2-8}

     & \multirow{2}{*}{\multirow{2}{*}\textbf{HuBERT-Large}\, \cite{hsu2021hubert}} & \multirow{2}{*}{\multirow{2}{*}\textbf{310M}} & \multirow{2}{*}{\multirow{2}{*}\textbf{LibriLight}} & No Aug.  & $\text{{{68.03}}}_{\pm 5.19}$ & $\text{{{42.18}}}_{\pm 2.35}$  & $\text{{{55.10}}}_{\pm 1.42}$ \\
     & & & & \emph{RepAugment} & $\text{{{67.75}}}_{\pm 3.21}$ & $\text{{{43.45}}}_{\pm 3.40}$ & $\text{{\textbf{55.60}}}_{\pm 0.38}$ \\
     \cmidrule{2-8}

     & \multirow{2}{*}{\multirow{2}{*}\textbf{XLS-R-300M}\, \cite{babu2021xls}} & \multirow{2}{*}{\multirow{2}{*}\textbf{310M}} & \multirow{2}{*}{\makecell{Multilingual\\LibriSpeech}} & No Aug.  & $\text{{{69.36}}}_{\pm 2.58}$ & $\text{{{43.23}}}_{\pm 2.06}$  & $\text{{{56.29}}}_{\pm 0.81}$ \\
     & & & & \emph{RepAugment} & $\text{{{68.62}}}_{\pm 6.53}$ & $\text{{{44.83}}}_{\pm 4.46}$ & $\text{{\textbf{56.73}}}_{\pm 1.19}$ \\

    \bottomrule

    \end{tabular}
    }
    \vspace{5mm}
\end{table*}

%% file: Tables/table2.tex
\begin{table*}[!t]
    \centering
    \caption{Overall performance on the ICBHI dataset for the official 60--40\% train--test split task of respiratory sound classification. 
    Here, IN, AS, and MLS refer to ImageNet~\cite{deng2009imagenet}, AudioSet~\cite{audioset}, and Multilingual LibriSpeech~\cite{pratap20_interspeech} together with other speech datasets as described in~\cite{babu2021xls}, respectively.
    \textbf{Best} and {\underline{second best}} are highlighted.}
    
    \label{tab:table2}
    \renewcommand{\arraystretch}{1}
    \addtolength{\tabcolsep}{8pt}
    \resizebox{\linewidth}{!}{
    \begin{tabular}{llcl|lll}
    \toprule
    Method & Model & Pretrain & Venue & $S_p$\,(\%) & $S_e$\,(\%) & \textbf{Score}\,(\%) \\
    \hline \midrule
    
    RespireNet \cite{gairola2021respirenet} (CBA+BRC+FT) & ResNet34 & IN & \textit{EMBC`21} & 72.30 & 40.10  & 56.20 \\
    
    Ren \textit{et al.} \cite{ren2022prototype} & CNN8-Pt & - & \textit{ICASSP`22} & 72.96 & 27.78 & 50.37 \\
    
    Chang \textit{et al.} \cite{chang22h_interspeech} & CNN8-dilated & - & \textit{INTERSPEECH`22} & 69.92 & 35.85 & 52.89 \\
    
    Wang \textit{et al.} \cite{wang2022domain} (Splice) & ResNeSt & IN & \textit{ICASSP`22} & 70.40 & 40.20 & 55.30 \\
    
    Moummad \textit{et al.} \cite{moummad2023pretraining}\,(SCL) & CNN6 & AS & \textit{WASPAA`23} & 75.95 & 39.15 & 57.55 \\    
    
    Bae \textit{et al.} \cite{bae23b_interspeech}\, (Fine-tuning)  & AST & IN\,+\,AS & \textit{INTERSPEECH`23} & $\text{77.14}$ & $\text{41.97}$ & $\text{59.55}$ \\
    
    Bae \textit{et al.} \cite{bae23b_interspeech}\, (Patch-Mix CL) & AST & IN\,+\,AS & \textit{INTERSPEECH`23} & $\text{\underline{81.66}}$ & $\text{{43.07}}$ & $\text{\bf 62.37}$ \\
    
    Kim \textit{et al.} (AFT on Mixed-500) \cite{kim2023adversarial}\, & AST & IN\,+\,AS & \textit{NeurIPSW`23} & $\text{{80.72}}$ & $\text{{42.86}}$ & $\text{\underline{61.79}}$ \\

    Kim \textit{et al.} (DAT) \cite{kim2023stethoscope}\, & AST & IN\,+\,AS & \textit{ICASSP`24} & $\text{77.11}$ & $\text{42.50}$ & $\text{59.81}$ \\
    
    Kim \textit{et al.} (SG-SCL) \cite{kim2023stethoscope}\, & AST & IN\,+\,AS & \textit{ICASSP`24} & $\text{{79.87}}$ & $\text{\underline{43.55}}$ & $\text{{61.71}}$ \\

    \midrule

    \textbf{RepAugment [ours]} & \text{XLS-R-300M} & MLS & - & $\text{{68.62}}_{\pm 6.53}$ & $\text{\bf 44.83}_{\pm 4.46}$ & $\text{56.73}_{\pm 1.19}$ \\
    
    \textbf{RepAugment [ours]} & AST & IN\,+\,AS & - & $\text{\textbf{82.47}}_{\pm 5.39}$ & $\text{40.55}_{\pm 4.70}$ & $\text{61.51}_{\pm 0.51}$ \\
    
    \bottomrule
    \end{tabular}}
\end{table*}

%% file: Tables/table3.tex
\begin{table*}[!t]
    \centering
    \caption{Accuracy (\%) of the under-represented classes on the ICBHI test set for AST and XLS-R-300M fine-tuning under different augmentation strategies.
    \textbf{Bold} denotes the best result.}\label{tab:table3}
    \addtolength{\tabcolsep}{7pt}
    \scriptsize{
    \resizebox{\linewidth}{!}{
    \begin{tabular}{p{2pt}lc|c|c|cc|c}
    \toprule
    & &  & \multicolumn{5}{c}{Data Augmentation Strategy} \\
    \cmidrule(l{2pt}r{2pt}){4-8}
    Model & Class & Ratio\,(\%) & No Aug. & SpecAugment & Rep-Mask & Rep-Gen & RepAugment\\
                                      \hline \midrule
    \multirow{4}{*}{AST} & normal ($c_{normal}$) & 57.29                          & $\text{76.82}_{\pm 2.47}$        & $\text{{75.52}}_{\pm 7.03}$         & $\text{77.71}_{\pm 6.38}$         & $\text{80.57}_{\pm 2.00}$  & $\text{\textbf{82.48}}_{\pm 5.39}$ \\

    & crackle ($c_{crackle}$) & 23.55                          & $\text{51.62}_{\pm 4.17}$        & $\text{\textbf{59.32}}_{\pm 3.75}$         & $\text{53.68}_{\pm 5.04}$         & $\text{51.96}_{\pm 3.79}$  & $\text{52.30}_{\pm 3.84}$ \\
    
    & wheeze ($c_{wheeze}$) & 13.97                           & $\text{35.56}_{\pm 2.87}$        & $\text{33.92}_{\pm 4.35}$         & $\text{\textbf{40.15}}_{\pm 3.17}$        & $\text{39.82}_{\pm 3.85}$  & $\text{40.08}_{\pm 4.84}$   \\
    
    & both ($c_{both}$) & 5.19                              & $\text{14.54}_{\pm 4.71}$         & $\text{18.46}_{\pm 4.73}$         & $\text{17.48}_{\pm 3.62}$        & $\text{20.14}_{\pm 4.93}$ & $\text{\textbf{21.68}}_{\pm 4.65}$ \\ 
    \bottomrule
    
    \multirow{4}{*}{XLS-R} & normal ($c_{normal}$) & 57.29                          & $\text{68.28}_{\pm 2.89}$        & -         & $\text{70.87}_{\pm 9.14}$         & $\text{68.08}_{\pm 1.02}$  & $\text{\textbf{70.89}}_{\pm 2.25}$ \\

    & crackle ($c_{crackle}$) & 23.55                          & $\text{52.81}_{\pm 2.76}$        & -         & $\text{54.27}_{\pm 8.68}$         & $\text{53.56}_{\pm 2.61}$  & $\text{\textbf{56.24}}_{\pm 5.14}$ \\

    & wheeze ($c_{wheeze}$) & 13.97                           & $\text{31.52}_{\pm 4.76}$        & -         & $\text{{31.64}}_{\pm 9.80}$        & $\text{\textbf{39.74}}_{\pm 6.45}$  & $\text{36.99}_{\pm 5.22}$   \\
    
    & both ($c_{both}$) & 5.19                              & $\text{18.32}_{\pm 7.97}$         & -         & $\text{15.94}_{\pm 3.26}$        & $\text{19.58}_{\pm 6.87}$ & $\text{\textbf{24.68}}_{\pm 8.28}$ \\ 
    \bottomrule
    
    \end{tabular}
    }}
\end{table*}

%% file: 5-conclusion.tex
\vspace{1mm}
\section{CONCLUSIONS}

In this paper, we explored the potential of pretrained speech models for respiratory sound classification. We find that there exists characterization gap between speech and respiratory sounds, which limits their efficacy. Further, compared to spectrogram-based approaches that also benefit greatly from SpecAugment, waveform-based approaches perform poorly. To alleviate the issue, we propose an input-agnostic data augmentation that operates on the feature representations of the network. We find that our approach is effective as data augmentation for waveform-based models and even outperforms SpecAugment in some cases.